# Polynomial Solution of Non-Central Potentials

Sameer M. Ikhdair* and Ramazan Sever†

*Department of Physics, Near East University, Nicosia, North

Cyprus, Mersin 10, Turkey

†Department of Physics, Middle East Technical University, 06531

Ankara, Turkey.

(March 9, 2018)

## Abstract

We show that the exact energy eigenvalues and eigenfunctions of the Schrödinger equation for charged particles moving in certain class of non-central potentials can be easily calculated analytically in a simple and elegant manner by using Nikiforov and Uvarov (NU) method. We discuss the generalized Coulomb and harmonic oscillator systems. We study the Hartmann Coulomb and the ring-shaped and compound Coulomb plus Aharanov-Bohm potentials as special cases. The results are in exact agreement with other methods.

*sikhdair@neu.edu.tr

†sever@metu.edu.tr

# I. INTRODUCTION

A large number of important physical problems require solving the non-relativistic and relativistic wave equations for charged particles moving in central and non-central potentials to determine the energy eigenvalues and the corresponding eigenfunctions. It is known that for very limited potentials, the Schrödinger equation is exactly solvable. In general, one has to resort to numerical techniques or approximation schemes. For many of the quantum mechanical systems, most popular approximation methods such as shifted $1/N$ expansion, WKB method, perturbation theory, path integral solution, supersymmetric quantum mechanics and the idea of shape invariance are widely used for this purpose. But some of these methods have drawbacks in application. Although some methods give simple relations for the eigenvalues, however, they give very complicated relations for the eigenfunctions.

The study of exact solutions of the Schrödinger equation for a class of non-central potentials with a vector potential and a non-central scalar potential is of considerable interest in quantum chemistry [1-10]. In recent years, numerous studies [11-15] have been made in analyzing the bound states of an electron in a Coulomb field with simultaneous presence of Aharanov-Bohm (AB) [16] field, and/or a magnetic Dirac monopole [17], and Aharanov-Bohm plus oscillator (ABO) systems. In most of these studies, the eigenvalues and eigenfunctions are obtained via seperation of variables in spherical or other orthogonal curvilinear coordinate systems. Recently, the path integral for particles moving in non-central potentials is evaluated to derive the energy spectrum of this system analytically [18]. Further, the idea of supersymmetry and shape invariance is also used to obtain exact solutions of such non-central but seperable potentials [19]. The radial and angular pieces of the Laplacian of the Schrödinger equation can be treated by using the idea of shape invariance within the same framework.

Very recently, an alternative method called as Nikiforov-Uvarov (NU) method, which received much interest, has been introduced for solving Schrödinger equation [20] for some well known analytically solvable potentials [21-26]. Similarly, Dirac, Klein-Gordon equations for Coulomb potential [27,28], generalized Hulthen [29,30], usual and modified Woods-Saxon potentials are also solved by using the NU method [26,31-33].

This paper is structured as follows: In Section II, we briefly introduce the basic concepts of the Nikiforov-Uvarov (NU) method. In Section III, we have obtained the eigenvalues and eigenfunctions of the generalized Coulomb system and generalized oscillatory potentials in the Schrödinger equation using the NU method. In Section IV, we give some concluding remarks.

## II. BASIC CONCEPTS OF THE METHOD

The solution of the Schrödinger-like second-order differential equations plays a vital role in studying many important problems of theoretical physics. In this regard, the Nikiforov and Uvarov (NU) method can be used to solve these types of equations with an appropriate coordinate transformation $s = s(r)$ [20]:

$$\psi_n''(s) + \frac{\widetilde{\tau}(s)}{\sigma(s)}\psi_n'(s) + \frac{\widetilde{\sigma}(s)}{\sigma^2(s)}\psi_n(s) = 0, \tag{1}$$

where $\sigma(s)$ and $\widetilde{\sigma}(s)$ are the polynomials with at most of second-degree, and $\widetilde{\tau}(s)$ is a first-degree polynomial. The special orthogonal polynomials [20] reduce Eq.(1) to a simple form by employing $\psi_n(s) = \phi_n(s)y_n(s)$, and choosing an appropriate function $\phi_n(s)$. Consequently, Eq.(1) can be reduced into an equation of the following hypergeometric type:

$$\sigma(s)y_n''(s) + \tau(s)y_n'(s) + \lambda y_n(s) = 0, \tag{2}$$

where $\tau(s) = \widetilde{\tau}(s) + 2\pi(s)$ (its derivative must be negative) and $\lambda$ is a constant given in the form

$$\lambda = \lambda_n = -n\tau'(s) - \frac{n(n-1)}{2}\sigma''(s), \qquad n = 0, 1, 2, ... \tag{3}$$

It is worthwhile to note that $\lambda$ or $\lambda_n$ are obtained from a particular solution of the form $y(s) = y_n(s)$ which is a polynomial of degree $n$. Further, $y_n(s)$ is the hypergeometric-type function whose polynomial solutions are given by Rodrigues relation

$$y_n(s) = \frac{B_n}{\rho(s)} \frac{d^n}{ds^n} \left[\sigma^n(s)\rho(s)\right], \tag{4}$$

where $B_n$ is the normalization constant and the weight function $\rho(s)$ must satisfy the condition [20]

$$\frac{d}{ds}w(s) = \frac{\tau(s)}{\sigma(s)}w(s), \ w(s) = \sigma(s)\rho(s). \tag{5}$$

In order to determine the weight function given in Eq.(5), we must obtain the following polynomial:

$$\pi(s) = \frac{\sigma'(s) - \widetilde{\tau}(s)}{2} \pm \sqrt{\left(\frac{\sigma'(s) - \widetilde{\tau}(s)}{2}\right)^2 - \widetilde{\sigma}(s) + k\sigma(s)}. \tag{6}$$

In principle, the expression under the square root sign in Eq.(6) can be arranged as the square of a polynomial. This is possible only if its discriminant is zero. In this case, an equation for $k$ is obtained. After solving this equation, the obtained values of $k$ are included in the NU method and here there is a relationship between $\lambda$ and $k$ by $k = \lambda - \pi'(s)$. After this point an appropriate $\phi_n(s)$ can be extracted from the condition

$$\frac{\phi'(s)}{\phi(s)} = \frac{\pi(s)}{\sigma(s)}. \tag{7}$$

### III. APPLICATIONS

Non-central potentials are normally not discussed in most text books on quantum mechanics. This is presumably because most of them are not analytically solvable. However, it is worth noting that there is a class of non-central potentials in three dimensions (3D) for which the Schrödinger equation is seperable. This section deals with two systems involving a generalized form of such potentials: a one known in quantum chemistry as the Hartmann system and another one of much use in quantum chemistry and nuclear physics. Both systems correspond to ring-shaped potentials; namely, the Coulomb-kepler and the isotropic harmonic oscillator system in 3D. Three-dimensional potentials that are singular along curves have received a great deal of attention in recent years. In particlar, the Coulombic ring-shaped potential [1-4] revived in quantum chemistry by Hartmann and coworkers [5,6] and the oscillatory ring-shaped potential [7-9] systematically studied by Quesne [10] have been investigated from a quantum mechanical point of view by using various approaches. As ring-shaped systems, they may play an important role in all situations. For example, the Coulombic-Kepler system is of interest for ring-shaped molecules like cyclic polyenes [5,6]. Further, the harmonic oscillator system is of potential use in the study of super-deformed nuclei.

In the following we solve the Schrödinger equation in 3D for its eigenvalues and eigenfunctions with non-central but seperable potentials of much intrest.

### A. The generalized Coulomb system

The Coulombic ring-shaped (Hartmann) potential [5,6] in spherical coordinates is

$$V(r,\theta) = \frac{A}{r} + \frac{B}{r^2 \sin^2\theta} + \frac{C\cos\theta}{r^2 \sin^2\theta}, \tag{8}$$

where $A = -Ze^2$, $B$ and $C$ are real constants. The potential in (8) introduced by Makarov et al. [34] and its importance lies on the fact that compound Coulomb plus Aharanov-Bohm potential [16] and Hartmann ring-shaped potential, originally proposed as model for the benzene molecule [5] are mathematically linked to this potential. In fact the energy spectrum for these two potentials can be obtained directly by considering these as special case of the general non-central seperable potential.

Our aim is to derive the energy spectrum for a moving charged particle in the presence of a potential given by Eq.(8) analytically in a very simple way. The Schrödinger equation in spherical polar coordinates for a particle in the presence of a potential $V(r,\theta)$ is

$$\left\{ -\frac{\hbar^2}{2\mu} \left[ \frac{1}{r^2} \frac{\partial}{\partial r} \left( r^2 \frac{\partial}{\partial r} \right) + \frac{1}{r^2} \left( \frac{1}{\sin\theta} \frac{\partial}{\partial \theta} \left( \sin\theta \frac{\partial}{\partial \theta} \right) + \frac{1}{\sin^2\theta} \frac{\partial^2}{\partial \varphi^2} \right) \right] \right.$$

$$\left. + V(r,\theta) - E_n \right\} \Psi_{nlm}(r,\theta,\varphi) = 0, \tag{9}$$

where $\mu = \frac{m_1 m_2}{m_1 + m_2}$ being the reduced mass and $\Psi_{nlm}(r, \theta, \varphi)$ being the total wave function defined by

$$\Psi_{nlm}(r, \theta, \varphi) = R(r)\Theta(\theta)e^{\pm im\varphi}, \quad R(r) = g(r)/r, \quad m = 0, 1, 2, \cdots. \tag{10}$$

The Schrödinger equation in Eq.(9) can be reduced to the two ordinary differential equations, one for an electron in a Coulomb-like field:

$$\frac{d^2 R(r)}{dr^2} + \frac{2}{r}\frac{dR}{dr} + \left[\frac{2m}{\hbar^2}\left(E_{nl} - \frac{A}{r}\right) - \frac{l(l+1)}{r^2}\right] R(r) = 0, \tag{11}$$

and one allows us to obtain the standard properties of the spherical harmonics given by the angular part:

$$\frac{d^2 \Theta(\theta)}{d\theta^2} + \frac{\cos\theta}{\sin\theta}\frac{d\Theta(\theta)}{d\theta} + \left[l(l+1) - \frac{(m^2 + B + C\cos\theta)}{\sin^2\theta}\right]\Theta(\theta) = 0. \tag{12}$$

We shall first solve the radial part of the Schrödinger equation. Hence, by using the transformation $s = r$, Eq.(11) can be transformed into the form given by Eq.(1) as

$$R''(s) + \frac{2}{s}R'(s) + \left[\frac{-\kappa_n^2 s^2 - \beta^2 s - l(l+1)}{s^2}\right] R(s) = 0, \tag{13}$$

where

$$\kappa_n^2 = -\frac{2mE_n}{\hbar^2} > 0 \ (E_n < 0), \quad \beta^2 = \frac{2mA}{\hbar^2}, \tag{14}$$

representing the bound-state case. Further, defining the following associated polynomials:

$$\widetilde{\tau}(s) = 2, \quad \sigma(s) = s, \quad \widetilde{\sigma}(s) = -\kappa_n^2 s^2 - \beta^2 s - l(l+1), \tag{15}$$

and using Eq.(6), $\pi(s)$ is found in four possible values as [26,30-32]

$$\pi(s) = \begin{cases} \kappa_n s + l, & k_1 = -\beta^2 + (2l+1)\kappa_n, \\ -(\kappa_n s + l + 1), & k_1 = -\beta^2 + (2l+1)\kappa_n, \\ \kappa_n s - l - 1, & k_2 = -\beta^2 - (2l+1)\kappa_n, \\ -\kappa_n s + l, & k_2 = -\beta^2 - (2l+1)\kappa_n, \end{cases} \tag{16}$$

where the roots are determined by means of the same procedures as in [26,30-32]. We also find

$$\tau(s) = \begin{cases} 2(\kappa_n s + l + 1), & k_1 = -\beta^2 + (2l+1)\kappa_n, \\ -2(\kappa_n s + l), & k_1 = -\beta^2 + (2l+1)\kappa_n, \\ 2(\kappa_n s - l), & k_2 = -\beta^2 - (2l+1)\kappa_n, \\ -2(\kappa_n s - l - 1), & k_2 = -\beta^2 - (2l+1)\kappa_n. \end{cases} \quad (17)$$

Imposing the condition $\tau'(s) < 0$, one gets the following physical particular solution: for the case $k_2 = -\beta^2 - (2l+1)\kappa_n$, we obtain $\lambda_n = 2n\kappa_n$ and $\lambda = -\beta^2 + 2(l+1)\kappa_n$. With $\lambda = \lambda_n$, the energy eigenvalues obtained from Eq.(11) for the Coulomb potential ($A = -Ze^2$) are

$$E_{n,l} = \frac{-2\mu(Ze^2)^2}{[2\hbar(n+l+1)]^2}, \quad n = 0, 1, 2, \cdots, \quad (18)$$

which is the desired bound-state satisfying the boundary conditions [35]. To find eigenfunctions, we first determine the weight function which satisfies the conditions in Eq.(15) and Eq.(17), using Eq.(5) we get

$$\rho(r) = e^{-2\kappa_n r} r^{2l+1}, \quad (0 \leq r < \infty). \quad (19)$$

Substituting Eq.(19) into Eq.(4), we obtain the first part of the wavefunction

$$y_n(s) = C_n L_n^{(2l+1)}(2\kappa_n r), \quad (20)$$

where $L_n(r)$ being the Laguerre Polynomials and $C_n$ being the normalization factor. Using Eq.(7), we find the other part of the wavefunction as

$$\phi_n(s) = r^l e^{-\kappa_n r}. \quad (21)$$

Finally, for bound states, the radial part of the wavefunction (finite for all $r$) has the form:

$$R_n(r) = N_n r^l e^{-\kappa_n r} L_n^{(2l+1)}(2\kappa_n r), \quad (0 \leq r < \infty), \quad (22)$$

where $N_n$ being a new normalization constant. It is clear from Eq.(22) that the radial wave function does vanish at the origin ($r = 0$) [35]. Hence, Eq.(18) and Eq.(22) are true bound state solutions [35].

Repeating for the angular part by using the simple transformation $s = \cos\theta$, one can rewrite Eq.(12) in the following form

$$\frac{d^2\Theta(\theta)}{ds^2} - \frac{2s}{1-s^2}\frac{d\Theta(\theta)}{ds} + \frac{1}{(1-s^2)^2}\left[-l(l+1)s^2 - Cs + l(l+1) - m^2 - B\right]\Theta(\theta) = 0. \quad (23)$$

Further, defining the following associated polynomials

$$\widetilde{\tau}(s) = -2s, \quad \sigma(s) = 1 - s^2, \quad \widetilde{\sigma}(s) = -l(l+1)s^2 - Cs - m^2 - B + l(l+1), \quad (24)$$

and using Eq.(6), $\pi(s)$ can be found in four possible values as

$$\pi(s) = \begin{cases} \sqrt{A_1}s + \sqrt{A_2}, & k_1 = l(l+1) - A_1, \\ -\sqrt{A_1}s + \sqrt{A_2}, & k_1 = l(l+1) - A_1, \\ \sqrt{A_2}s + \sqrt{A_1}, & k_2 = l(l+1) - A_2, \\ -\sqrt{A_2}s + \sqrt{A_1}, & k_2 = l(l+1) - A_2, \end{cases} \quad (25)$$

where $A_1 = \left(\frac{m^2+B-\sqrt{(m^2+B)^2-C^2}}{2}\right)$ and $A_2 = \left(\frac{m^2+B+\sqrt{(m^2+B)^2-C^2}}{2}\right)$ being strictly positive constants. The roots are determined by means of the same previous procedures as before. We also find $\tau(s)$ as

$$\tau(s) = \begin{cases} 2\left[\left(\sqrt{A_1}-1\right)s + \sqrt{A_2}\right], & k_1 = l(l+1) - A_1, \\ -2\left[\left(\sqrt{A_1}+1\right)s - \sqrt{A_2}\right], & k_1 = l(l+1) - A_1, \\ 2\left[\left(\sqrt{A_2}-1\right)s + \sqrt{A_1}\right], & k_2 = l(l+1) - A_2, \\ -2\left[\left(\sqrt{A_2}+1\right)s - \sqrt{A_1}\right], & k_2 = l(l+1) - A_2. \end{cases} \quad (26)$$

Imposing the condition $\tau'(s) < 0$, we also get the following two particular solutions:

i) For the case $k_1 = l(l+1) - A_1$, one obtains $\lambda = l(l+1) - A_1 - \sqrt{A_1}$ and $\lambda_{\widetilde{n}} = \widetilde{n}(\widetilde{n}+1) + 2\widetilde{n}\sqrt{A_1}$, which consequently give the following quantum number:

$$l = \widetilde{n} + \sqrt{A_1}, \quad (27)$$

where $\widetilde{n}$ is a non-negative integer.

Following similar procedures in finding the radial part of the wave function, we also find the angular part as

$$\Theta(\theta) = (1-\cos\theta)^{(\sqrt{A_1}-\sqrt{A_2})/2}(1+\cos\theta)^{(\sqrt{A_1}+\sqrt{A_2})/2} P_{nl}^{(\sqrt{A_1}-\sqrt{A_2},\sqrt{A_1}+\sqrt{A_2})}(\theta),\ \Phi(\varphi) = e^{\pm i\sqrt{A_1}\varphi},$$

(28)

with $\sqrt{A_1} = -n + l; -n - l - 1$.

ii) For the case $k_2 = l(l+1) - A_2$, one obtains $\lambda = l(l+1) - A_2 - \sqrt{A_2}$ and $\lambda_{\tilde{n}} = \tilde{n}(\tilde{n}+1) + 2\tilde{n}\sqrt{A_2}$, which consequently give the following quantum number:

$$l = \tilde{n} + \sqrt{A_2},$$

(29)

where $\tilde{n}$ is a non-negative integer. Moreover, the angular part of the wavefunction reads

$$\Theta(\theta) = (1-\cos\theta)^{(\sqrt{A_2}-\sqrt{A_1})/2}(1+\cos\theta)^{(\sqrt{A_2}+\sqrt{A_1})/2} P_{nl}^{(\sqrt{A_2}-\sqrt{A_1},\sqrt{A_2}+\sqrt{A_1})}(\theta),\ \Phi(\varphi) = e^{\pm i\sqrt{A_2}\varphi},$$

(30)

with $\sqrt{A_2} = -n + l; -n - l - 1$. For the case $(B = C = 0,\ A_1 = 0,\ A_2 = m^2)$, Eq.(30) becomes

$$\Theta(\theta) = \sin^m\theta P_{nl}^{(m,m)}(\cos\theta),\ \Phi(\varphi) = e^{\pm im\varphi},$$

(31)

with $m = -n + l; -n - l - 1$.

Therefore, our final energy levels and eigenfunctions for a real bound charged particle in a Coulombic field potential plus a combination of non-central potentials given by Eq.(12) are

$$E_{n,\tilde{n},m} = \frac{-\mu(Ze^2)^2}{2\hbar^2\left[n+\tilde{n}+1+\left(\frac{m^2+B+\sqrt{(m^2+B)^2-C^2}}{2}\right)^{1/2}\right]^2},$$

(32)

and

$$\Psi_{n,\tilde{n},m}(r,\theta,\varphi) = N_n r^l s^{-\kappa_n r} L_n^{(2l+1)}(2\kappa_n r)(1-\cos\theta)^{(\sqrt{A_2}-\sqrt{A_1})/2}$$
$$\times (1+\cos\theta)^{(\sqrt{A_2}+\sqrt{A_1})/2} P_{nl}^{(\sqrt{A_2}-\sqrt{A_1},\sqrt{A_2}+\sqrt{A_1})}(\theta) e^{\pm i\sqrt{A_2}\varphi},$$

(33)

respectively, with the angular quantum number satisfying the condition given by Eq.(29). These results are consistent with that in Ref.[19]. Further, for $C = 0$, it agrees with the corresponding energy spectrum already been calculated by the path integral solution of the system [18] and using KS transformation [36,37]

### 1. Hartmann ring-shaped potential

The Hartmann ring-shaped potential in question is of the form [38]

$$V(r,\theta) = \eta\sigma^2 \left( \frac{2a_0}{r} - \frac{\eta a_0^2}{r^2 \sin^2\theta} \right) \varepsilon_0, \tag{34}$$

where $a_0 = \frac{\hbar^2}{\mu e^2}$ and $\varepsilon_0 = -\frac{\mu e^4}{2\hbar^2}$, are the Bohr's radius and H-atom ground state energy, respectively. The dimensionless parameters $\eta$ and $\sigma$ are positive and real which range from about 1 to 10 in molecular applications. This potential had been proposed by Hartmann as a model for molecules like benzene [5] and can be obtained from the non-central potential of Eq.(8) after setting $Z = \eta\sigma^2$, $B = \eta^2\sigma^2$ and $C = 0$. Therefore, the energy spectrum and wavefunctions, for this system can be written immediately from Eq.(32) and Eq.(33) as

$$E_{n,\widetilde{n},m} = \frac{-\mu(\eta\sigma^2 e^2)^2}{2\hbar^2 \left[ n + \widetilde{n} + 1 + \sqrt{m^2 + (\eta\sigma)^2} \right]^2}, \tag{35}$$

and

$$\Psi_{n,\widetilde{n},m}(r,\theta,\varphi) = N_n r^l s^{-\epsilon_n r} L_n^{(2l+1)}(2\epsilon_n r) \sin^{\sqrt{m^2+\eta^2\sigma^2}}(\theta) P_{nl}^{(\sqrt{m^2+\eta^2\sigma^2},\sqrt{m^2+\eta^2\sigma^2})}(\theta) e^{\pm i\sqrt{m^2+\eta^2\sigma^2}\varphi}, \tag{36}$$

respectively, with $l = \widetilde{n} + \sqrt{m^2 + (\eta\sigma)^2}$, $\widetilde{n} = 0,1,2,\cdots$. This result agrees with that of in Refs.[34,38] where energy spectrum has been calculated by solving Schrödinger equation using the path integral solution.

### 2. Aharonov-Bohm plus Dirac monopole potential

We consider the general case involving a vector potential $\mathbf{A} = \frac{F(\theta)}{r\sin\theta}\widehat{e}_\varphi$, with $F(\theta) = \Phi/2\pi + g(1 - \cos\theta)$, where $\Phi$ is the magnetic flux, together with a scalar potential $V(r) = -\frac{Ze^2}{r}$. This problem of a relativistic Dirac electron in the presence of a Coulombic field plus Dirac magnetic monopole and the Aharonov-Bohm (AB) potentials have been discussed by Hoang et al. [39]. Both potentials can be chosen to be of arbitrary strength depending on the values of the coupling constants $\Phi$ and $g$. The effective potential for this system is

$$V_{ee}(r,\theta) = -\frac{Ze^2}{r} + \left[\frac{a-b}{r^2 \sin^2\theta}\right], \qquad (37)$$

where $a = e^2 F^2(\theta)$ and $b = 2meF(\theta)$. This potential is a special case of the non-central potential of Eq.(8) with $A = -Ze^2$, $B = a - b$, and $C = 0$. Now if we define

$$q = -ge, \quad \widetilde{m} = \frac{e\Phi}{2\pi} - q - m, \qquad (38)$$

this gives

$$B = (\widetilde{m} + q\cos\theta)^2 - m^2. \qquad (39)$$

Therefore, Eq.(23) may be rewritten as

$$\frac{d^2\Theta(\theta)}{ds^2} - \frac{2s}{1-s^2}\frac{d\Theta(\theta)}{ds} + \frac{1}{(1-s^2)^2}\left[-\left[q^2 + l(l+1)\right]s^2 - 2q\widetilde{m}s + l(l+1) - \widetilde{m}^2\right]\Theta(\theta) = 0. \qquad (40)$$

By defining the following associated polynomials

$$\widetilde{\tau}(s) = -2s, \quad \sigma(s) = 1 - s^2, \quad \widetilde{\sigma}(s) = -\left[q^2 + l(l+1)\right]s^2 - 2q\widetilde{m}s + l(l+1) - \widetilde{m}^2, \qquad (41)$$

and using Eq.(6), $\pi(s)$ can also be found in four possible values as

$$\pi(s) = \begin{cases} (|\widetilde{m}|s + |q|), & k_1 = l(l+1) + q^2 - \widetilde{m}^2, \text{ if } |q| < |\widetilde{m}|, \\ -(|\widetilde{m}|s + |q|), & k_1 = l(l+1) + q^2 - \widetilde{m}^2, \text{ if } |q| < |\widetilde{m}|, \\ (|q|s + |\widetilde{m}|), & k_2 = l(l+1), \text{ if } |\widetilde{m}| < |q|, \\ -(|q|s + |\widetilde{m}|), & k_2 = l(l+1), \text{ if } |\widetilde{m}| < |q|, \end{cases} \qquad (42)$$

where the roots are determined by means of the same previous procedures. The $\tau(s)$ is obtained as

$$\tau(s) = \begin{cases} 2\left[(|\widetilde{m}| - 1)s + |q|\right], & k_1 = l(l+1) + q^2 - \widetilde{m}^2, \text{ if } |q| < |\widetilde{m}|, \\ -2\left[(|\widetilde{m}| + 1)s + |q|\right], & k_1 = l(l+1) + q^2 - \widetilde{m}^2, \text{ if } |q| < |\widetilde{m}|, \\ 2\left[(|q| - 1)s + |\widetilde{m}|\right], & k_2 = l(l+1), \text{ if } |\widetilde{m}| < |q|, \\ -2\left[(|q| + 1)s + |\widetilde{m}|\right], & k_2 = l(l+1), \text{ if } |\widetilde{m}| < |q|. \end{cases} \qquad (43)$$

Imposing the condition $\tau'(s) < 0$, we get two particular solutions:

i) For the case $k_1 = l(l+1) + q^2 - \widetilde{m}^2$, we obtain $\lambda = l(l+1) + q^2 - \widetilde{m}^2 - |\widetilde{m}|$ and $\lambda_{\widetilde{n}} = \widetilde{n}(\widetilde{n}+1) + 2\widetilde{n}|\widetilde{m}|$, which in turn give the following quantum number:

$$l = -\frac{1}{2} + \sqrt{\left[\widetilde{n} + |\widetilde{m}| + \frac{1}{2}\right]^2 - q^2},\ \widetilde{n} = 0, 1, 2, \cdots. \tag{44}$$

and hence energy spectrum, Eq.(18), with the previous quantum number becomes

$$E_{n,\widetilde{n}} = \frac{-\mu(Ze^2)^2}{2\hbar^2 \left[n + \frac{1}{2} + \sqrt{\left(\widetilde{n} + |\widetilde{m}| + \frac{1}{2}\right)^2 - q^2}\right]^2},\ |q| < |\widetilde{m}|, \tag{45}$$

and the wavefunction is given in Eq.(28) with the allowed quantum numbers defined by Eq.(44).

ii) For the case $k_2 = l(l+1)$, we obtain $\lambda = l(l+1) - |q|$ and $\lambda_{\widetilde{n}} = \widetilde{n}(\widetilde{n}+1) + 2\widetilde{n}|q|$, which in turn give the following quantum number:

$$l = -\frac{1}{2} + \sqrt{\left[\widetilde{n} + |q| + \frac{1}{2}\right]^2 - q^2},\ \widetilde{n} = 0, 1, 2, \cdots. \tag{46}$$

Hence, the energy spectrum, Eq.(18), with the previous quantum number in Eq.(46), becomes

$$E_{n,\widetilde{n}} = \frac{-\mu(Ze^2)^2}{2\hbar^2 \left[n + \frac{1}{2} + \sqrt{\left(\widetilde{n} + |q| + \frac{1}{2}\right)^2 - q^2}\right]^2},\ |q| > |\widetilde{m}|. \tag{47}$$

Further, the wavefunction is given in Eq.(30) with the allowed quantum numbers defined by Eq.(46).

### B. A generalized Oscillatory potential

The well-known ring-shape oscillator potential

$$V(r,\theta) = A^2 r^2 + \frac{B}{r^2 \sin^2 \theta}, \tag{48}$$

where $B$ being strictly positive constant. The limiting case $C = 0$ corresponds to an isotropic harmonic oscillator and will serve as a testing for the results to be obtained.

The Schrödinger equation given by Eq.(9), for a particle in a ring-shape oscillator potential can be reduced to an ordinary differential equation in the radial part:

$$\frac{d^2 g(r)}{dr^2} + \frac{2\mu}{\hbar^2}\left[E_{nl} - A^2 r^2 - \frac{l(l+1)\hbar^2}{2\mu r^2}\right] g(r) = 0, \quad g(r) = rR(r), \tag{49}$$

and another one for the angular part:

$$\frac{d^2 \Theta(\theta)}{d\theta^2} + \frac{\cos\theta}{\sin\theta}\frac{d\Theta(\theta)}{d\theta} + \left[l(l+1) - \frac{(m^2 + B)}{\sin^2\theta}\right]\Theta(\theta) = 0. \tag{50}$$

We shall first solve the radial part of the Schrödinger equation. Employing a simple transformation $s = r^2$, we get

$$g''(s) + \frac{1}{2s}g'(s) + \left[\frac{-\alpha^2 s^2 + \varepsilon_n^2 s - l(l+1)}{4s^2}\right]g(s) = 0, \tag{51}$$

with

$$\varepsilon_n^2 = \frac{2\mu E_n}{\hbar^2} \quad (E_n > 0), \quad \alpha^2 = \frac{2\mu A^2}{\hbar^2}. \tag{52}$$

Further, defining the following associated polynomials:

$$\tilde{\tau}(s) = 1, \quad \sigma(s) = 2s, \quad \tilde{\sigma}(s) = -\alpha^2 s^2 + \varepsilon_n^2 s - l(l+1), \tag{53}$$

and using Eq.(6), $\pi(s)$ can also be found in four possible values

$$\pi(s) = \begin{cases} \alpha s + l + 1, & k_1 = \frac{\varepsilon_n^2}{2} + \alpha(l + \frac{1}{2}), \\ -(\alpha s + l), & k_1 = \frac{\varepsilon_n^2}{2} + \alpha(l + \frac{1}{2}), \\ \alpha s - l, & k_2 = \frac{\varepsilon_n^2}{2} - \alpha(l + \frac{1}{2}), \\ -\alpha s + l + 1, & k_2 = \frac{\varepsilon_n^2}{2} - \alpha(l + \frac{1}{2}), \end{cases} \tag{54}$$

where $k_{1,2}$ is determined by means of same procedures as before. The $\tau(s)$ is also obtained as

$$\tau(s) = \begin{cases} 3 + 2\alpha s + 2l, & k_1 = \frac{\varepsilon_n^2}{2} + \alpha(l + \frac{1}{2}), \\ 1 - 2\alpha s - 2l, & k_1 = \frac{\varepsilon_n^2}{2} + \alpha(l + \frac{1}{2}), \\ 1 + 2\alpha s - 2l, & k_2 = \frac{\varepsilon_n^2}{2} - \alpha(l + \frac{1}{2}), \\ 3 - 2\alpha s + 2l, & k_2 = \frac{\varepsilon_n^2}{2} - \alpha(l + \frac{1}{2}). \end{cases} \tag{55}$$

Imposing the condition $\tau'(s) < 0$, one gets the following physical solution: For the case $k_2 = \frac{\varepsilon_n^2}{2} - \alpha(l+\frac{1}{2})$, we obtain have $\lambda_n = 2\alpha n$ and $\lambda = \frac{\varepsilon_n^2}{2} - \alpha(l+\frac{3}{2})$. With $\lambda = \lambda_n$, the energy eigenvalues obtained are [40]:

$$E_{nl} = \sqrt{\frac{\hbar^2}{2\mu}(4n+2l+3)A}, \ n = 0, 1, 2, \cdots. \tag{56}$$

Further, to find eigenfunctions, we first determine the weight function which satisfies the conditions in Eq.(53) and Eq.(55), using Eq.(5) we get

$$\rho(r) = e^{-\alpha r^2} r^{(2l+1)}, \ (0 \leq r < \infty). \tag{57}$$

Substituting Eq.(57) into Eq.(4), we obtain the first part of the wavefunction

$$y_n(r) = C_n L_n^{(l+1/2)}(\alpha r^2), \tag{58}$$

where $C_n$ being the normalization factor. Further, using Eq.(7), the other part of the wavefunction is

$$\phi_n(r) = r^{l+1} e^{-\alpha r^2/2}. \tag{59}$$

Finally, the radial part of the wavefunction has the general form:

$$R_{nl}(r) = N_n r^l e^{-\alpha r^2/2} L_n^{-(l+1/2)}(\alpha r^2), \tag{60}$$

where $N_n$ is a new normalization factor. It is clear from Eq.(60) that this case is a true bound state solution [35]. The total wavefunction can be obtained from the combined solutions of Eq.(30) after setting $C = 0$ and also from Eq.(60) as

$$\Psi_{nlm}(\theta, \varphi) = A_n r^l e^{-\alpha r^2/2} L_n^{-(l+1/2)}(\alpha r^2) \sin^{\sqrt{m^2+B}}(\theta) P_{nl}^{(\sqrt{m^2+B},\sqrt{m^2+B})}(\theta) e^{\pm i\sqrt{m^2+B}\varphi}, \tag{61}$$

where $A_n$ being a new normalization constant.

## IV. RESULTS AND CONCLUSIONS

In this work, the Schrödinger equation with a class of non-central but seperable potentials has been studied. The analytical expressions for the energy levels and eigenfunctions have also been found analytically for every case. This method is simple and useful in solving other complicated systems analytically. Therefore, it is found that the energy eigenvalues are consistent with the results obtained by using other methods. We also point out that these exact results obtained for the non-central potentials may have some interesting applications in the study of different quantum mechanical systems and atomic physics.

## ACKNOWLEDGMENTS

S.M. Ikhdair wishes to dedicate this work to his family for their love and assistance.